# Current-induced Pinwheel Oscillations in Perpendicular Magnetic Anisotropy Spin Valve Nanopillars

Richard Choi[1], J.A. Katine[2], Stephane Mangin[3], and Eric E. Fullerton[1]

[1]Center for Memory Recording Research, University of California San Diego, La Jolla, CA 92093-0401 USA
[2]HGST San Jose Research Center, San Jose, CA 95135 USA
[3]Institut Jean Lamour, Université de Lorraine, CNRS UMR 7198 B.P. 70239, F-54506, Vandoeuvre, France

**Nanopillar spin valve devices are typically comprised of two ferromagnetic layers: a reference layer and a free layer whose magnetic orientation can be changed by both an external magnetic field and through the introduction of spin-polarized electric current. Here we report the continuous repeated switching behavior of both the reference and free layers of a perpendicular spin valve made of Co/Pd and Co/Ni multilayers that arises for sufficiently large DC currents. This periodic switching of the two layers produces an oscillating signal in the MHz regime but is only observed for one sign of the applied current. The observed behavior agrees well with micromagnetic simulations.**

## I. Introduction

The use of giant magnetoresistance (GMR), tunneling magnetoresistance (TMR) and related spin-torque switching devices and their ability to exploit electrical control over magnetization dynamics has sparked extensive development for applications in data recording, magnetic memory and spin torque nano-oscillators [1-4]. The basic phenomena of spin torque occur for current flowing through two magnetic layers separated by a non-magnetic spacer where the current becomes spin polarized by transmission through or upon reflection from the first magnetic layer (the reference layer) and interacts with the second ferromagnetic layer (the free layer). The spin-torque interaction with the free layer can lead to deterministic switching or high-frequency oscillations of the free layer.

In most devices the reference layer is designed to be stable in response to spin-torque interactions so that the switching behavior of the device depends solely on the action on the free layer. However, for sufficiently high currents the interactions with the free layer can destabilize the reference layer [5]. More generally one expects that the mutual interaction between the layers should cause the two layer to rotate in the same direction in a pinwheel-like motion as originally predicted in [6] and discussed for layers with in-plane magnetic anisotropy [7]. A similar behavior may be expected for spin valve devices with perpendicular magnetic anisotropy (PMA) [8] as shown schematically in Fig. 1 where each magnetic layer is considered as a macrospin. Assuming that you start with a parallel alignment (Fig. 1a) and positive DC currents (negative electron flow) the spin torque interactions will stabilize the reference layer and destabilize the free layer (as a result of the back flow of down spin electrons reflected from the reference layer) causing the free layer to switch to the configuration shown in Fig. 1b. This is the standard operation of a memory cell where current is used to set the direction of the free layer. In this configuration the spin-torque interaction now stabilizes the free layer but will tend to destabilize the reference layer which, for sufficiently high currents, should then switch to Fig. 1c. In memory cell applications the current is below that needed to perturb the reference layer. In the configuration of Fig. 1c the spin-torque interaction will again destabilize the free layer causing it to switch to the configuration in Fig. 1d. Finally the spin-torque interaction will cause the reference layer to again switch to Fig. 1e which is the starting configuration (Fig. 1a) and the process repeats.

As the layers reverse the system oscillates between the parallel and antiparallel configurations causing a periodic change in the resistance. This behavior should be, in principle, independent of the current flow direction. In the present study we observe, for the first time, the pinwheel motion in magnetic spin valves with PMA as shown schematically in Fig. 1 [3], [4] where the resistance oscillates in a periodic fashion. For typical spin torque nano oscillators the oscillations result from precessional motion of the magnetic layer giving oscillation frequencies in the GHz range [9]. However, pinwheel motion requires periodic reversal of the direction of the magnetic layers which is typically on the nanosecond or longer time scales [10]. Thus the expected oscillation frequencies are in the MHz range.

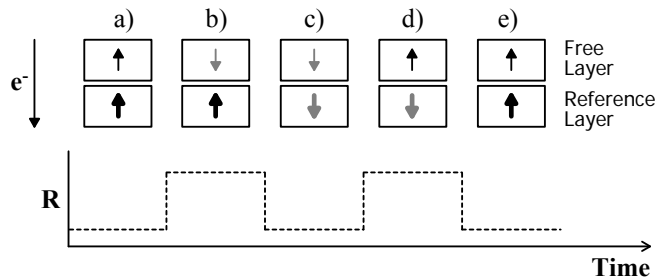

Fig. 1. (a)-(e) Schematic of expected magnetic configuration and device resistance versus time for large positive DC current (negative electron flow as shown in Figure). The top layer is the lower anisotropy free layer and the bottom layer is the higher anisotropy reference layer, the arrows represent macrospin orientations.

## II. NANOPILLAR

The experimentally studied devices are identical in structure to those in Ref. [5], [Co(0.3nm)/Pd(0.7nm)]x2, [Co(0.15nm)/Ni(0.6nm)]x2, Co(0.3nm), Cu(4nm), [Co(0.15nm)/Ni(0.6nm)]x2, [Co(0.3nm)/Pd(0.7nm)]. Both reference and free layers have PMA due to the Co-Ni and Co-Pd multilayer structure [11]. The reference layer has higher anisotropy and is magnetically harder due to the additional repeats of the Co-Pd multilayer. The films were grown by dc magnetron sputtering onto Si(100) wafers coated with a 200-nm thermal oxide. The nanopillars were patterned by e-beam lithography into various geometries. In this study we focused on 120-nm diameter circular nanopillars.

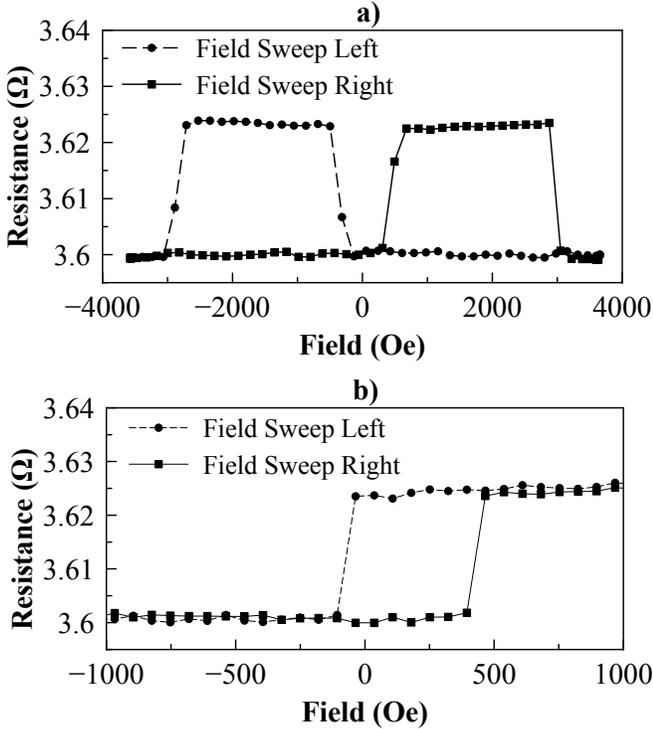

Fig. 2. (a) The field-dependent switching characteristics of the 120-nm circular nanopillars. The reference layer exhibits a coercive field of 3000 Oe, while the free layer (b) exhibits a coercive field of 250 Oe with an effective dipolar field of 180 Oe and GMR of approximately 0.62%.

## III. MEASUREMENTS

The electrical characterization was carried out using an AC lock-in amplifier with the current flowing perpendicular to the plane. We also applied an external magnetic field normal to the layers. In a 4-wire measurement, these devices exhibited a GMR ratio ranging from 0.5%-1.1%. The coercive fields were ~250 Oe and ~3000 Oe for the free and reference layers, respectively (Fig. 2) and the effective dipolar field acting on the free layer from the reference layer is 180 Oe as reflected in the shift of the minor loop in Fig. 2b.

We determined the minimum current to perturb the reference layer direction by using the approached outlined in Ref. [5] where both layers are set to the P state, we apply a large current for 1 s and then measure the response of the spin valve at low fields to see if there was a change in the reference layer. From these measurements the *I-H* switching boundaries of the reference layers for the nanopillar can be determined. Comparing these results to previous measurements we know that the critical current for free layer switching is much lower than for the reference layer [8].

Due to the spin torque efficiency dependence with respect to direction of current in metallic systems [12], one current direction has greater efficiency in switching the reference layer as shown in Fig. 3. For electrons flowing from the free layer to the reference layer we see a higher efficiency (as reflected in the increased slope in of *I-H* boundary) in reference layer switching than for electrons flowing in the opposite direction. In contrast to the conventional operation of a spin valve, current from the free layer is switching the reference layer state, this is an essential condition for oscillation to occur in these devices.

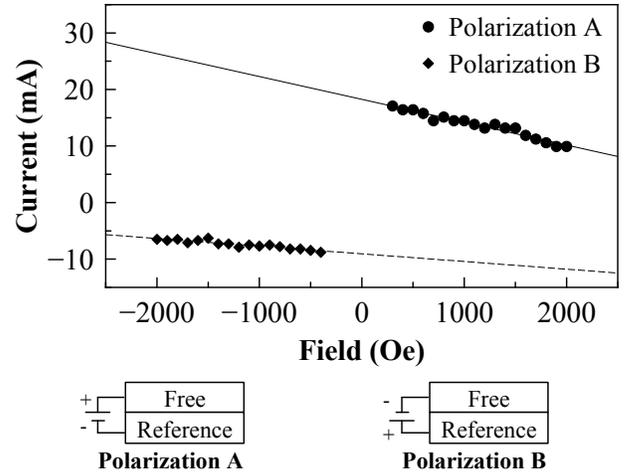

Fig. 3. The switching boundary and switching efficiency of the reference layer with respect to the direction of current. Polarization A represents the current direction where the free layer is positively biased, polarization B represents the current direction where the reference layer is positively biased.

High frequency measurements were performed by applying a DC current of 10-30 mA corresponding to a current density on the order of 1x10^8 A/cm^2 and a magnetic field perpendicular to the plane of the sample. While the current remains constant, the rapidly changing magnetic state produces a high frequency output which we record using a spectrum analyzer through a bias-tee.

## IV. RESULTS

The oscillation can only occur when the current is above the threshold value necessary to switch the reference layer (Fig. 3). High frequency activity can be observed for applied current up to the Joule heating limit of the nanopillar device. The characteristics and behaviors are similar among devices of identical geometry and show a demonstrable and repeatable tunability in peak frequency with respect to current (Fig. 4) where the characteristic frequency shifts to higher frequency with increasing current. While the peaks shift linearly, there is a widening of the bandwidth at higher currents due to less uniformity in the reversal behavior (as described below in the modeling).

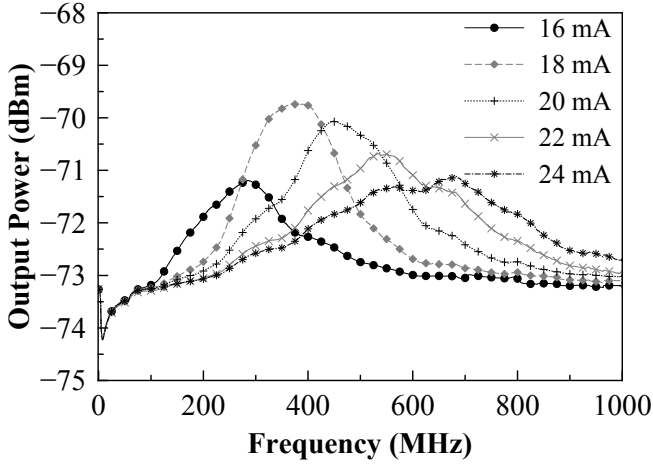

Fig. 4. Measured frequency output from oscillating nanopillar devices with a 1111 Oe bias field. Only DC current sources were used.

Unlike with current, the switching frequency of the nanopillar is not strongly affected by the magnitude of the applied magnetic field as shown in Fig. 5. Because each magnetic layer switches both direction as seen in Fig. 1 the applied field will speed up the switching into the field direction and slow the switching in the direction opposing the applied field yielding a frequency that is relatively insensitive to the applied field. The magnitude of the signal, however, increases with increasing magnetic field. As will be detailed in the modeling below the presence of a magnetic field increases the magnitude of the output signal by stabilizing the free layer in the field direction to increase the disparity in magnetization with respect to the reference layer, thereby increasing the change in resistance.

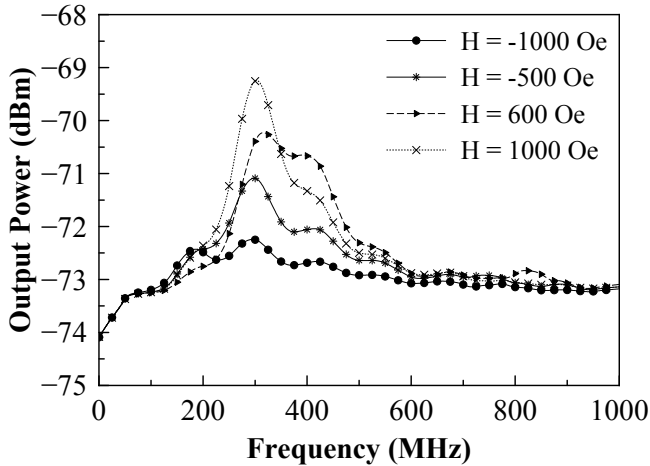

Fig. 5. Changing the field does little to alter the frequency at which oscillations occur, however higher field intensity can effectively help pin the free layer so larger GMR signals are produced, resulting in higher output power. A 17 mA DC current was applied to the device.

Unlike in simple model in Fig. 1 the oscillation behavior depends strongly on the direction of the applied current. We only observe oscillations when current polarized by the free layer flows towards the reference layer, polarization B as shown in Fig. 3. This may be expected as the reference layer is the harder layer to switch and the spin-torque efficiency for switching the reference layer is highest for polarization B (Fig. 3). We will further discuss that stable domains states are also observed in modeling that suppress the oscillatory behavior for one current direction.

We can estimate the expected oscillatory frequency from Refs. [13] and [14] where we know that the switching time $\tau$ of the magnetic layers follow the form

$$\frac{1}{\tau} = A(I - I_{c0}) \qquad (1)$$

where $A$ is the dynamic parameter which governs the switching rate, I is current and $I_{c0}$ is the zero temperature critical current. Equation (1) has been shown experimentally to describe the reversal of spin-torque devices with PMA layers [14]. The value of $I_{c0}$ can be estimated for both the free and reference layers

$$I_{c0} \approx \frac{\alpha M_s V}{g(\theta) p}(H_{Keff} - H) \qquad (2)$$

where $\alpha$ is the damping, $M_s$ is the saturation magnetization, $V$ the volume of the magnetic elements, $g(\theta)$ is expression for the angular variation in spin torque, $p$ is the spin polarization of current and ($H_{Keff} - H$) is the effective magnetic field acting magnetic layer where $H_{Keff}$ is the effective anisotropy field (including the demagnetization fields) of either the free and fixed layers and H is the external magnetic field that includes the dipolar fields originating from the adjacent magnetic layer.

The critical current for switching of both the reference and free layers in Fig. 3 follows a linear $I$ vs. $H$ behavior as expected from (2) and has been seen before [8]. Extrapolating the measured results in Fig. 6b to zero frequency give an estimated measure for $I_{c0}$ = 9.8 mA which agrees with the measured critical current results in Fig. 3 for the reference layers. However, because the measurements are at room temperature, the results in Fig. 3 slightly underestimate $I_{c0}$ because thermal energy will assist switching resulting in a measured critical current lower than predicted by Eq. (2) for longer times [10].

Since $I_{c0}$ for the reference layer is much larger than the free layer, the switching dynamics of the reference layer have a greater influence on the speed at which the system oscillates. With the frequency inversely proportional to the time constant $\tau$, Eq. (1) predicts the linear relationship between current and frequency we see in measurements.

While the applied bias field may affect signal intensity and noise, the effect does not significantly change the fundamental speed at which the device switches as shown in Fig. 5. Notably, $I_{c0}$ in Eq. (2) changes as the applied field changes, causing $\tau$ to change, while we may initially expect this to significantly impact the frequency of oscillation, we have in fact four switching events, the P-AP and AP-P switching for both reference and free layer $\tau_{f\_p\text{-}ap}$, $\tau_{f\_ap\text{-}p}$, $\tau_{rp\text{-}ap}$, $\tau_{r\_ap\text{-}p}$. For varying low values of field, as the P-AP switching time increases, the AP-P time decreases and vice versa resulting in very little change in the fundamental oscillating frequency when $H \ll H_{K\text{eff}}$.

## V. SIMULATION DATA

To further understand the pinwheel oscillations we have performed detailed micromagnetic simulations using parameters derived from characterization of these films. A 120-nm circular disc nanopillar was simulated with the FastMag simulation software [15] as well as the LLG micromagnetic simulator using the parameters presented in Table 1. A 1000-Oe field is applied along the direction perpendicular to the film plane. The AC response is presented in Fig. 6a and the transient response in Figs. 7a-b.

TABLE I
NANOPILLAR SIMULATION PARAMETERS

|  | Reference Layer | Free Layer |
|---|---|---|
| $K_1$ (erg/cm$^3$) | 3.0e6 | 2.1e6 |
| $M_s$ (emu/cm$^3$) | 600 | 600 |
| $A_{ex}$ (erg/cm) | 1.05e-6 | 1.05e-6 |
| $\alpha$ | 0.03 | 0.03 |
| Polarization | 0.35 | 0.35 |
| Magnetoresistance | 0.01 | 0.01 |

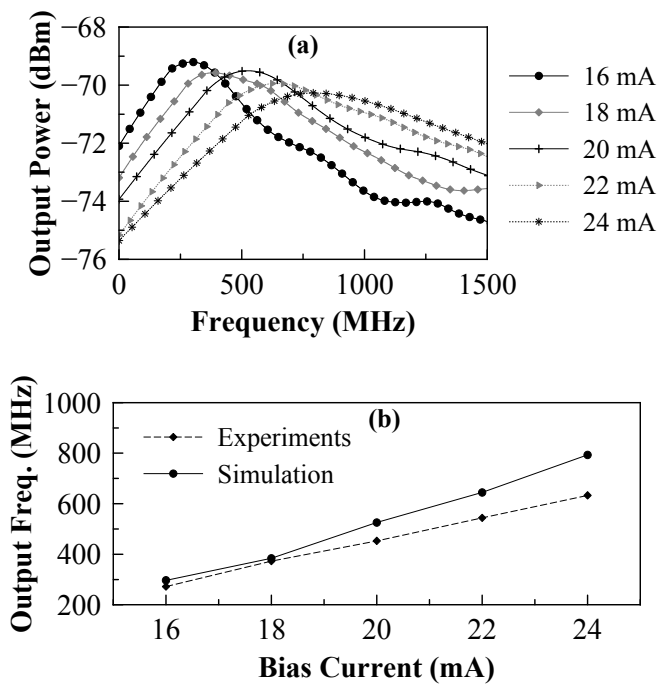

Fig. 6. (a) The simulated output spectrum for a 120nm nanopillar disc with 1000 Oe bias field and variable DC current. (b) The output frequency of the nanopillar versus input bias current, the behavior from the model closely resembles the characteristics of physical nanopillar.

Simulation results show behavior that closely matches the response of the measured devices as shown in Fig. 6b. We speculate that the slight deviation between the experimental and simulated oscillation frequencies are likely from deviations between input parameters and actual device characteristics. Additionally, these are zero temperature simulations where joule heating is also not calculated. The transient analysis in Fig. 7a-d portrays the output characteristics from the interactive switching behavior between the free and reference layer. The switching behavior is characterized by inhomogeneous magnetic configurations in each of the layers. This inhomogeneous switching in PMA devices has been well established from the presence of domain state in the free layer [16, 17] and direct imaging of the nonuniform reversal by time-resolved x-ray microscopy [18]. Domain states have also been observed in reversal of the reference layers by spin-torque switching [5]. We similarly observe non-universal reversal of both the free and reference layers in our simulations.

In the model the layers begins to switch when the opposite layer reaches a sufficient magnetization in the opposite orientation. Because of this we never observe the full uniform states shown schematically in Fig. 1. Both layers are able to begin switching before either layer is fully saturated, resulting in only a percentage of the device magnetoresistance contributing to the RF signal (Fig. 7b,c). In Fig. 8 we show snapshots of specific magnetic configurations. The application of an external applied field helps to achieve a uniform state, particularly in the free layer, increasing the output voltage. This non-uniform response results in the broad frequency response at a given current and relatively low Q. Thus the schematic shown in Fig. 1 is an oversimplification.

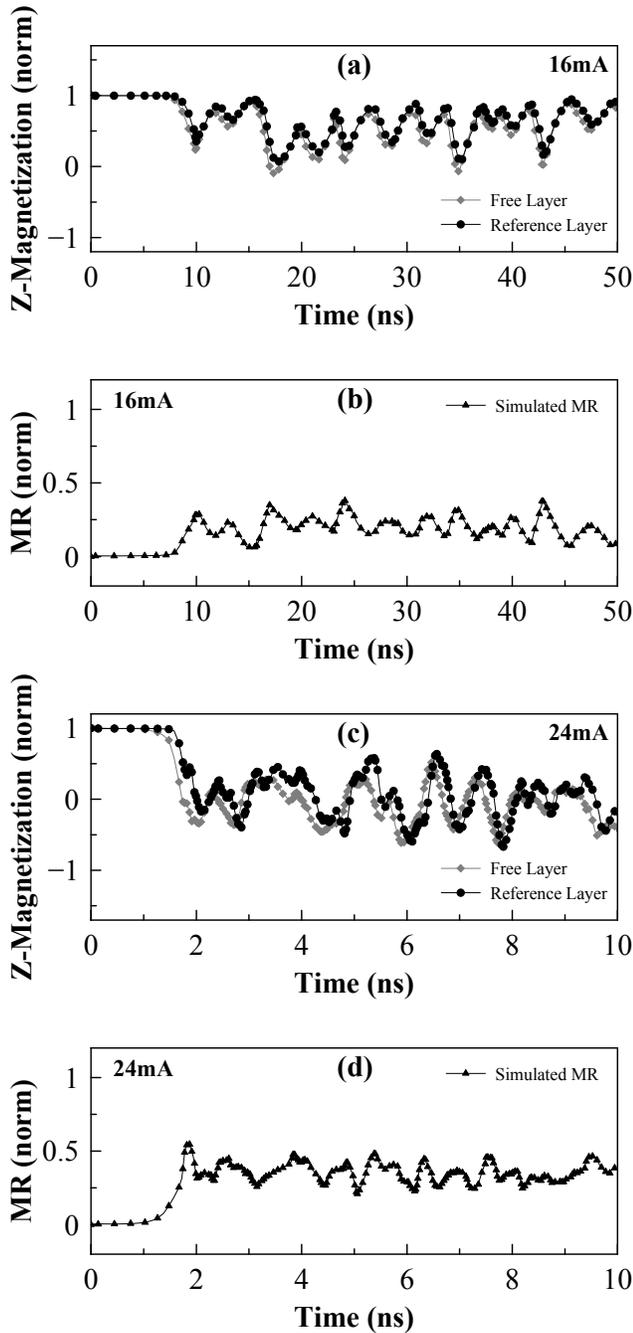

Fig. 7. The simulated behavior of the layers during oscillation with 16mA and 24 mA. (a) The transient response of both reference and free layers with 16 mA current. Oscillation frequency is more heavily determined by the switching characteristics of the reference layer, as the lower coercivity free layer mainly follows the reference layer as it switches. (b) The normalized spin resistance of the nanopillar with 16 mA as a function of time. As expected from the observation of the individual layers, only a fraction of the full GMR signal is reached. (c) As I increases, there is less difference between the relative switching time between the free and reference layers as seen with 24 mA of current. (d) The MR response at 24 mA illustrates the behavior where there is less coherence in the oscillation.

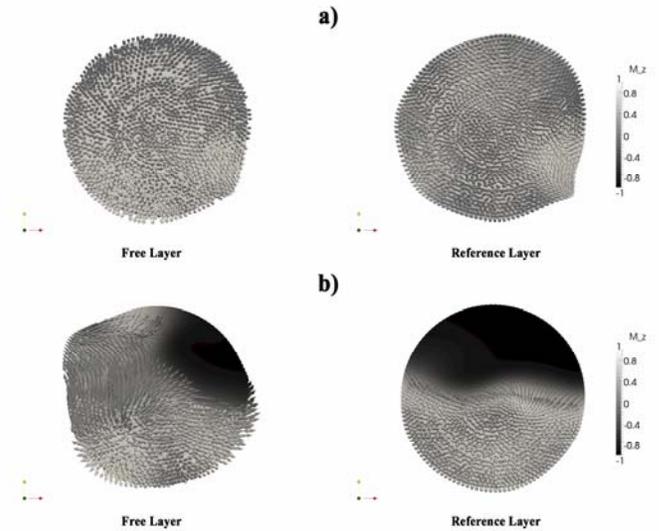

Fig. 8. Simulation data visualized in both relative P and AP states with identical axis orientation on both layers. This visualization illustrates the nature of the magnetic configuration during oscillation. Neither layer is able to have its moment fully polarized within the layers, thus these are not absolute P or AP state as shown in Fig. 1, but rather (a) relative P and (b) relative AP states.

The simulations further show that the system can lock into a stable, non-oscillating state. This was observed with the current applied in the opposite polarization as Fig. 7 but for otherwise identical conditions. This state has a parallel alignment of the magnetization between the free and reference layers. This stable state may explain why no oscillation signal is observed for this current direction in physical devices. However the simulations do suggest with sufficient current, the device may oscillate even in this current direction, although it exceeds the joule heating limit of the devices.

## VI. SUMMARY

PMA nanopillar spin valves exhibit tunable oscillations from repeated and high speed switching of both free and reference layers. It is observe that these devices are able to switch at a frequency in the microwave regime using only DC sources, and that the overall switching behavior is dominated by the characteristics of the reference layer. For current in the reverse direction these devices tend to lock into a stable state.

ACKNOWLEDGEMENTS

The authors would like to thank the Vitaliy Lomakin computational electromagnetics and micromagnetics group for access to the FastMag micromagnetics simulator. Majd Kuteifan, Marco Lubarda and Iana Volvach for their assistance in running the FastMag software. The research is supported by the Qualcomm Fellow-Mentor-Advisor (FMA) Program and the NSF award DMR #1312750. As well as ANR-NSF Project, ANR-13-IS04-0008-01 "COMAG."